# Terahertz brightness at the extreme: demonstration of 5 GV/m low frequency $\lambda^3$ terahertz bullet

Mostafa Shalaby[1] and Christoph P. Hauri[1,2]

[1]Paul Scherrer Institute, SwissFEL, 5232 Villigen PSI, Switzerland

[2]Ecole Polytechnique Federale de Lausanne, 1015 Lausanne, Switzerland

The brightness of a light source defines its applicability to nonlinear phenomena in science. Bright low frequency terahertz (< 5THz) radiation confined to a diffraction-limited spot size is a present hurdle due to the broad bandwidth and long wavelengths associated with single-cycle terahertz pulses as well as due to the lack of terahertz wavefront correctors. Here, using a present-technology system, we employ a new concept of terahertz wavefront manipulation and focusing optimization. We demonstrate a spatio-temporal confinement of terahertz energy at its physical limits to the least possible 3-dimensional light bullet volume of lambda cubic. This leads to a new regime of extremely bright terahertz radiation reaching 40 $PW/m^2$ intensity. The presented work is focused on the sub-5 THz range using small aperture organic crystals DSTMS and OH1. The obtained peak field of up to 5 GV/m and 17 Tesla is order of magnitude higher than any reported single-cycle field oscillation in the entire THz range from a laser-based system and surpassing large scale accelerator systems. The presented results are foreseen to have a great impact on future nonlinear terahertz applications in different science disciplines.

Request for material and correspondence: most.shalaby@gmail.com; christoph.hauri@psi.ch

Brightness defines a light source's ability to nonlinearly interact with matter by driving it out of an equilibrium state. Qualitatively, the brightness reveals the power concentrated at a given frequency in a solid angle according to $P / \lambda^2$ where P and $\lambda$ are the total power and wavelength. Heisenberg's uncertainty principle defines a fundamental limit for a source brightness as it correlates the minimum achievable spot size to a given $\lambda$, known as the diffraction limit [1-2]. To approach this ultimate limit experimentally, an excellent source beam quality and carefully designed beam imaging optics are required. In laser physics, the quality factor $M^2$ is commonly used to describe how much the beam quality of monochromatic radiation deteriorates from the ideal diffraction-limited case. A high quality beam ($M^2 \approx 1$) of ultrabroadband radiation is hard to achieve even for mature laser technology at near infrared wavelengths.

At terahertz (THz) frequencies (0.1-10 THz), beam quality and brightness are peculiar hurdles for several reasons. First, the THz generation efficiency and the resulting pulse power are low. Second, high brightness is by nature harder to achieve for THz than it is for radiation at shorter wavelengths. For example, for a given power, the maximum brightness of a single-cycle laser pulse at 1 THz is five orders of magnitude lower than its equivalent at a typical near infrared (NIR) frequency (300 THz) exploiting the $1 / \lambda^2$ dependence. Finally, beam imaging and focusing of multi-octave spanning THz pulses are demanding and less-developed than for other wavelength ranges [3].

Recently, the advance of intense THz sources towards $\approx 0.1$ GV/cm field strength enabled first observations of nonlinear light matter interactions [4-7], the induction of insulator-to-metal transitions [8] and the cause of DNA damage [9]. However, a large range of THz applications, ranging from ultrafast domain switching [10-12] to particle acceleration [13], requires a giant leap in brightness to match the theoretically predicted needs in field strength being on the order of a few GV/m. Intense THz radiation is mainly generated by employing electron accelerators [14-16] and laser-based systems. Although the former has been ahead in terms of intensity, the availability and versatility of lasers are dominating time-resolved THz high-field science. The most prominent techniques for intense THz pulse generation, by employing a femtosecond amplified laser pulse, are based on laser-driven ion acceleration [17], emission from a gas plasma [18], and optical rectification in inorganic (e.g. $LiNbO_3$) [19] and organic nonlinear crystals (e.g. OH1, DAST, DSTMS) [20-22]. So far, the peak fields from these systems have been limited to 0.12-0.5

GV/m, respectively. We mention that difference frequency generation was also used to generate intense radiation near the THz band. However, as the carrier frequency is >>10 THz, it is common sense to call it mid infrared (MIR) rather than THz radiation [23]. Compact laser-based $LiNbO_3$ THz sources deliver currently the largest pulse energy (125 μJ, [24]) but it needs a non-collinear pump configuration with sophisticated pulse front tilting to achieve phase-matching. The maximum reported field using $LiNbO_3$ was demonstrated at significantly lower pulse energy (3 μJ) only. This is a direct consequence of shortened THz beam quality induced by such a non-collinear pumping scheme which makes diffraction-limited focusing a challenge. Much brighter radiation is expected from collinearly pumped THz schemes as the THz wavefront is untilted and the intensity profile is symmetric and Gaussian.

In this report, using a conventional and collinear THz generation scheme based on optical rectification in the small-size, highly efficient organic crystals DSTMS and OH1, we show that the THz source brightness is dramatically enhanced by optimizing the pump wavefront radius of curvature. This leads to a close to perfect THz wavefront and consequently excellent THz focusing conditions with demonstrated extremely high field strengths. We show that the produced THz light bullet is confined to the $\lambda^3$ volume and significantly alters beam temporal and spectral characteristics along the propagation axis.

For our THz source we employ a 100 Hz Ti:sapphire driven 3-stage optical parametric amplifier (OPA) system with pulse duration of 65 ± 5 fs. This source is used to pump small-size organic crystals DSTMS and OH1 at 1.5 μm and 1.35 μm wavelengths. In our scheme an all-reflective telescope assembly is used to adapt the spherical wavefront curvature of the pump beam at the position of the THz emitting organic crystal where the pump fluence is kept constant. The organic crystals are excellently suited for Terahertz generation since they offer (i) phase-matched THz generation in the 1 - 5 THz range (where no inorganic crystal was proven efficient), (ii) a very high second-order nonlinear optical susceptibility (214 pm/V [25] and 240 pm/V [26]) and (iii) high damage threshold ~ 20 $mJ/cm^2$. More details on the system are given in the Supplementary material.

While efficient THz energy extraction from organic crystals has previously been reported [20,21], the advance of the present work is the introduction of a novel scheme for enhanced THz wavefront control and improved THz beam transport, which compensates for wavefront aberrations originating from the source and from THz imaging optics. This results in a dramatic increase of the THz brightness and intensity. Several challenges are associated with the collinear THz generation scheme commonly used for organic crystals. Low frequency components diverge faster and may get spatially filtered on a finite size focusing mirror. This in turn leads to distortion of the THz beam wavefront. In order to get the best beam quality in the THz focal region we introduce the following optimization procedure. (1) Optimizing the pump wavefront curvature in order to achieve the optimal THz wavefront, followed by (2) fast expansion of the THz beam to employ large f-number optics. The latter is done using a 1:4 telescope based on 1" and 2"-diameter off-axis parabolic mirrors. The former is done by adapting the spherical component of the pump wavefront. In order to reach the $\lambda^3$ light bullet we systematically adjusted the pump beam wavefront curvature and the THz generation plane (crystal position) to optimize the THz peak field and spot size. This surprisingly simple approach is used to reach the physical limits of THz focusing. It is worth mentioning that the pump wavefront corrections do not compensate only for the divergence of the THz beam, but also for the low-order wavefront aberration of the subsequent focusing system.

The temporal traces of the generated pulses retrieved by air biased coherent detection (ABCD) [27] in the focal waist are shown in Fig. 1a for DSTMS and OH1 with peak field reaching 5 GV/m (16.7 T). In the lambda-cubic regime presented here we observe that the measured single-cycle carrier oscillates faster than previously reported from an identical DSTMS [20] (with similar pump pulse duration and crystal thickness). Furthermore we observe a high-frequency amplitude modulation in the temporal trace of OH1. These observations are a direct consequence of the improved beam quality and tight focusing scheme where high frequency THz components get dominant in the beam waist at the diffraction limit. Such high frequency features were observed even with electro optical sampling (EOS) detection as we show in the Supplementary material. The spectral amplitudes are given in Fig. 1b showing THz emission between 0.1 to 12 THz. More details on the spectral features are given in the Supplementary material. We measured the THz spot size (Fig. 1c & 1d) using an uncooled pyroelectric array detector (NEC Inc., 23.5 μm pixel size). After optimization of wavefront and focusing, the average radii at $1/e^2$ were 93 μm and 70 μm from DSTMS and OH1. This is strikingly small under consideration of the spectra from these crystals. Thanks to wavefront control, the THz spot size is more than three times smaller than in previous reports (300 μm [20]) using the same crystal

DSTMS and similar input beam size. This tight focusing enhances the THz peak intensity by an order of magnitude, thanks to the optimized wavefront.

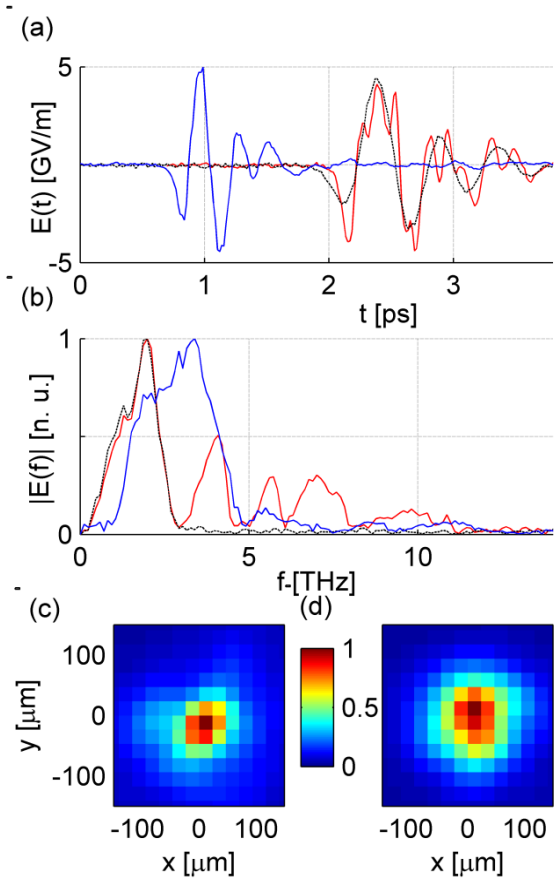

**Figure 1: The characteristics of the generated pulses.** (a) The temporal profiles of the THz pulses generated from DSTMS (blue) and OH1 (red and shifted in time). The peak fields reach 5 GV/m (16.7 T) and 4.4 GV/m (14.7 T) from the two crystals, respectively. The black curve shows the corresponding temporal profile from OH1 after removing the high frequency modulations using a 3 THz low pass filter. (b) The corresponding THz amplitude spectra. (c) The THz spot at the detector from (c) OH1 and (d) DSTMS, measured with a THz camera and shown in normalized units.

At lambda-cubic focusing the temporal and spectral shape of the THz are expected to alter rapidly across the waist as the Rayleigh length is minimized. In Fig. 2a & 2b, we show time domain maps of the THz temporal profile from DSTMS and OH1, respectively, reconstructed in the propagation ($z$) direction around the focus position ($z = z_0$). It illustrates the Gouy phase shift of $\pi$ across $(z - z_0) >> |z_R|$ with $z_R$ being the Rayleigh length. Fig. 2c & 2d, illustrates the temporal field evolution at positions $(z - z_0) = \{-4, 0, 4\}$ mm and the fast decay of field strength for $|z| > 0$ The spectral evolution across the waist (Fig. 2e, 2f, 2g

& 2h) shows a strong shift of the center frequency towards higher frequencies as we approach the focus position. This is a direct consequence of the $\lambda^3$ focusing scheme where the linear dependence of the spot size $\omega_0$ on the THz wavelength is highly pronounced as the $\lambda^3$ volume for lower frequencies is larger. These results contrast a recent report for THz focusing in (loose) non-$\lambda^3$ conditions. There, the spectrum is experimentally found to get broader around the focus with no shift in the center [20].

Evaluation of the beam quality parameter $M^2$ requires 3-dimensional beam profiling around the focus. As the THz radiation covers several octaves, the spectrum is added as a fourth dimension. We did time domain spectroscopy along the beam propagation direction across the focus using EOS detection. Then, we evaluated $z_R$ from the corresponding decrease in the axial beam intensity (Fig. 3a). From this, we obtained both the spot size (Fig. 3b) and an *effective* numerical aperture (Fig. 3c). This technique neglects the effect of the detector and assumes a highly symmetric Gaussian beam. In our case, it overestimates $z_R$ because of the overall large probe size that is nearly collimated along the scanned beam waist. Our experimental estimation of the effective NA was then compared against the NA calculated assuming a high quality diffraction-limited ($M^2 = 1$) beam. The good matching suggests that our beam has $M^2 \approx 1$ within a slight frequency dependence arising from the minimized-but-unavoidable divergence and imperfections in both the optical pump and generating crystals.

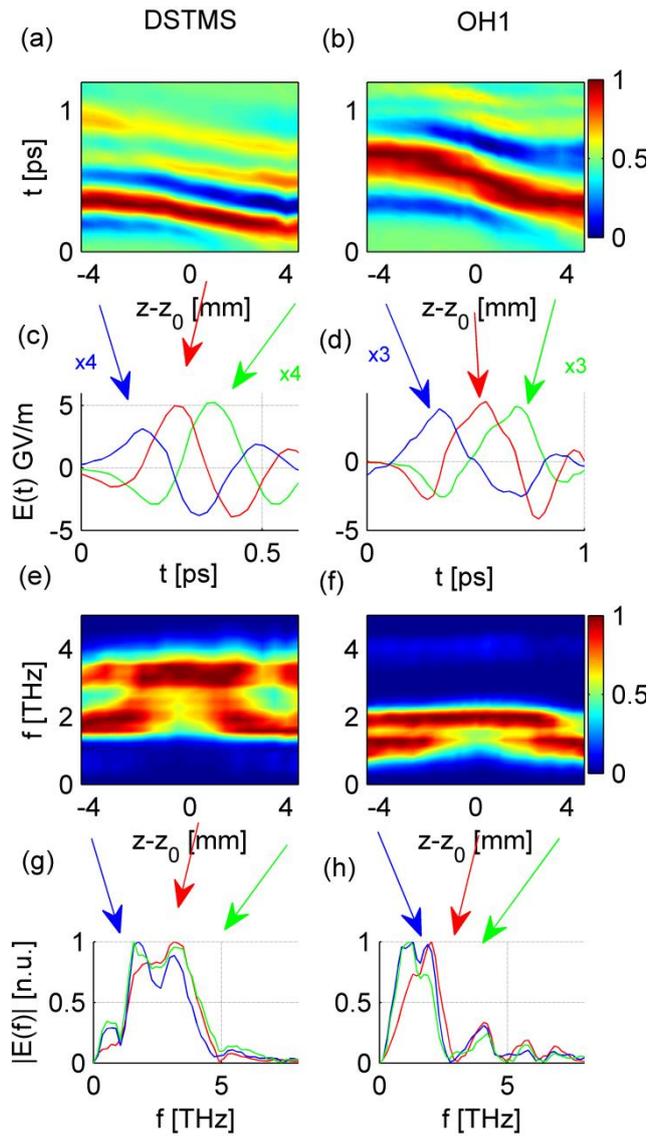

**Figure 2: Temporal and spectral modulation around the focus.** Two dimensional (normalized) maps of the THz time traces across the focus in the case of (a) DSTMS and (b) OH1, constructed with EOS detection. (c) & (d) Visualization of the Guoy effect across the focus with traces taken from (a) & (b). (e) & (f) Two dimensional maps of the normalized THz spectra versus position showing the shift of the spectrum center across the THz focus. $x_0$ is the position of the focal plane. (g) & (h) show the corresponding spectra of the time traces in (c) and (d).

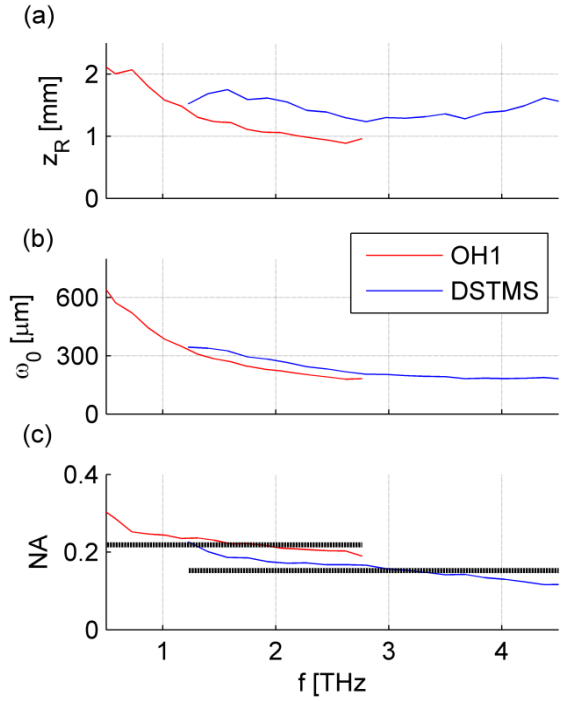

**Figure 3: Beam quality and diffraction-limited focusing.** (a) Rayleigh length, (b) minimum focus size, and (c) the effective numerical aperture estimated from the spot size. The solid dashed lines show the calculated numerical aperture assuming a diffraction-limited system with quality factor $M^2 = 1$.

In lambda-cubic focusing scheme the focus volume depends strongly on the THz frequency. Figure 4 illustrates this lambda-cubic dependence of the intensity in the focus. A set of LPF is used to measure the evolution of the spot size with frequency. As expected there is a linear decrease of the spot size for higher THz components. This is a consequence of the tight focusing where the intensities of the high frequency components go up dramatically overruling the contribution of the low frequency components to the overall imaged spot size. While the two crystals show a similar trend for radiation < 2 THz, its evolution with frequency is quite different (Fig. 4a & 4b). In the case of DSTMS, $\omega_0$ drops significantly for cut offs at 3 THz and 6 THz, followed by bare change for a further increase in the cut off frequency. This implies that the majority of the energy is concentrated at low frequencies (Fig. 4c), which is in line with the ABCD measurements shown in Fig. 1. In contrast, OH1 shows a much broader distribution of energy, especially in the 6-9 THz range. We measured the corresponding energy using a calibrated Golay cell (which has almost frequency independent response in our spectral range). The above-predicted energy buildup *rate* with frequency from the two crystals is well-demonstrated in Fig. 4d. The total energy measured was 75 μJ from

DSTMS and 52 µJ from OH1. The corresponding conversion efficiencies are 2.27 % and 1.46 %, respectively.

We measured peak electric fields of 5 GV/m (16.7 T) and 4.4 GV/m (14.7 T) from DSTMS and OH1, respectively (Fig. 4e). These values were obtained from small aperture organic crystals, in comparison with 0.5 GV/m obtained from a large aperture DSTMS [21]. Figure 4e shows the obtained peak fields when LPF with different cut off frequencies were used. Higher frequency components correspond to short pulse duration and, in our case, to small spot size as well. This, in turns, leads to higher field intensities (Fig. 4e). We estimated the peak fluence and intensity to be {480, 610} mJ/cm$^2$ and {33,26} PW/m$^2$ for DSTMS and OH1, respectively (Fig. 4f). Finally, the presented results considered the whole spectral range up to 18 THz. However, from DSTMS most of the energy is concentrated in the 1-5 THz, which we consider the most important part of this work as this range is hard to access. In the sub-5 THz range high peak field (3.2 GV/m, 10.7 T), peak fluence (226 mJ/cm$^2$) and peak intensities (13.6 PW/m$^2$) are achieved using DSTMS which are the highest ever reported values in the whole THz band so far. At the extreme peak fields reported here, even THz-induced ionization and damage could occur. Combining such intensities with field enhancement structures [8, 28-29] will lead to another giant leap in intensities.

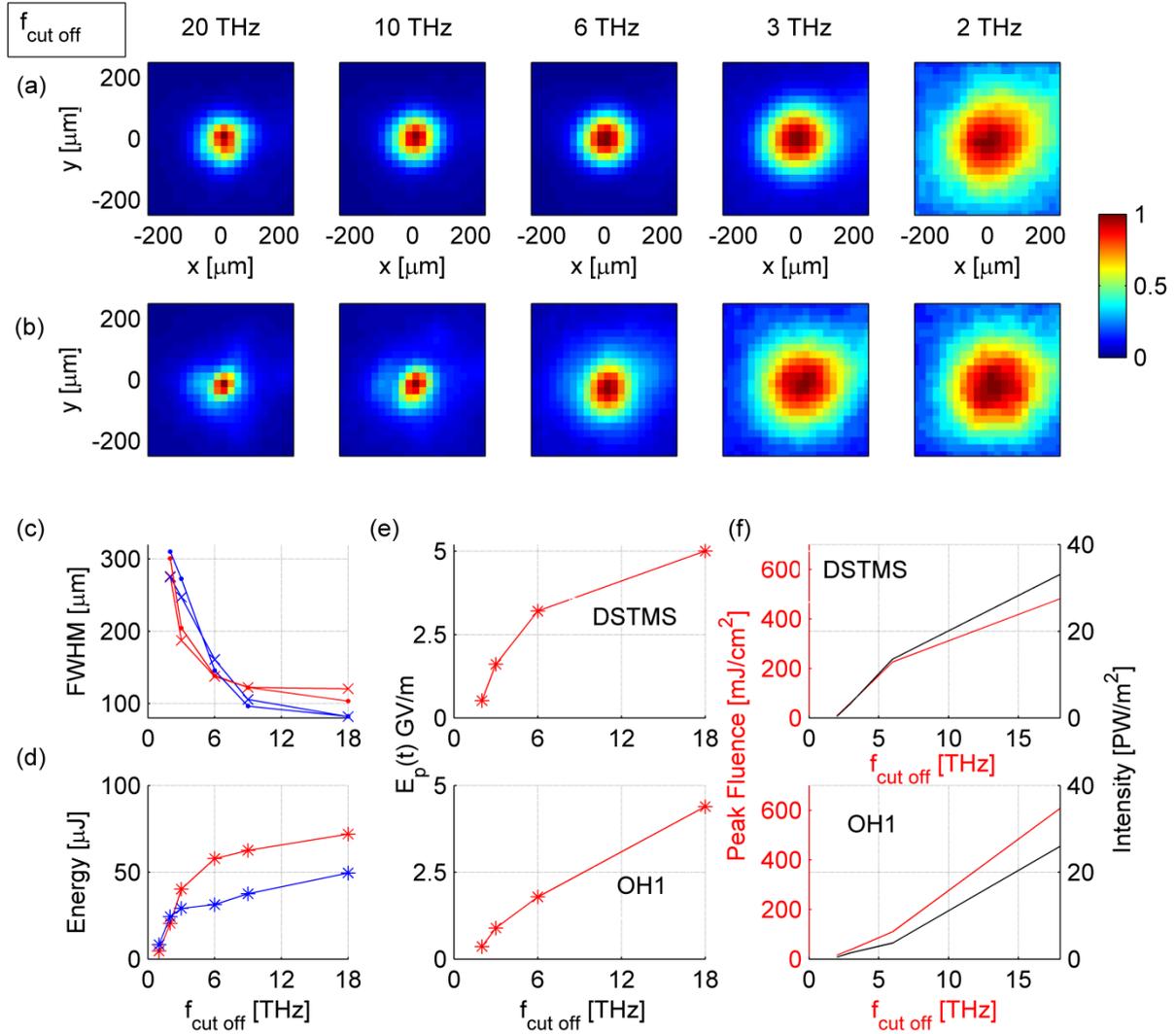

**Figure 4: Spot size, field and intensity characterization.** The normalized THz spot at the focus is shown for (a) DSTMS and (b) OH1 for different cut off frequencies {2,3,6,9,18} THz. (c) the extracted THz spot size FWHM from (a) & (b) for different cut off frequencies. Red and blue refer to DSTMS and OH1, respectively. The dot and cross markers denote the x and y directions. (d) The corresponding energies for DSTMS (red) and OH1 (blue). (e) The THz peak electric fields. (f) The measured peak fluence and peak intensities.

In conclusion, we have experimentally presented a $\lambda^3$ terahertz bullet using low frequency THz pulses. Our result is perhaps the first demonstration of such a system in the entire electromagnetic spectrum. This approach allowed us to reach, by far, the most intense low-frequency THz pulses between 1-5 THz

(DSTMS) and up to 18 THz (OH1). Our work introduces a new concept based on pump wavefront control to significantly enhance the THz beam quality and THz brightness in a collinear pump scheme based on a small-scale organic crystal. The method leads to extremely intense THz radiation of 33 PW/m$^2$ and field strength of 5 GV/m and 17 T. Such a compact ultra-intense THz source with such an extreme brightness will open up new avenues for nonlinear THz applications in a wide range of science.

**Acknowledgements** We are grateful to Marta Divall and Alexandre Trisorio for operating the Ti:sapphire laser system. The OPA operation was supported by Carlo Vicario and Marta Divall. The data acquisition software, motors control, and terahertz camera used in this work were previously installed by Carlo Vicario, Clemens Ruchert, Rasmus Ischebeck, Edwin Divall, and Balazs Monoszlai. We acknowledge financial support from the Swiss National Science Foundation (SNSF) (200021_146769). CPH acknowledges association to NCCR-MUST and support from SNSF grant no PP00P2_128493 and PP00P2_150732

# Supplementary material

**Terahertz generation system.** Our THz source consists of a 100 Hz Ti-Sapphire driven 3-stage Optical Parametric Amplifier (OPA) system with pulse duration around 65± 5 fs. This source is used to pump small-size organic crystals at 1.5 µm for DSTMS (thickness of 440 µm, diameter of 6 mm) and 1.35 µm for OH1 (thickness of 480 µm, diameter of 10 mm). The overall OPA conversion efficiency is ~ 39%. Taking beam transmission losses into account, pump energies of 3.3 mJ and 3.5 mJ are available for pumping the nonlinear crystal, at the two respective wavelengths. The measurements presented here have been done at the maximum energies. The spot size of the pump beam on the crystal had 1/e$^2$ dimensions of 3.8 mm and 4.1 mm, respectively, corresponding to a peak fluence of ~ 23 mJ/cm$^2$. This is close to the crystals' damage threshold. The crystal cut and orientation was chosen for the maximum THz generation, following ref [25-26].

Downstream the organic crystal, the THz beam was expanded by an all-reflective telescope and then focused on the detector using off-axis mirrors. To attenuate the THz beam, we used a set of 400 µm-thick Silicon wafers (each having almost frequency-independent transmission amplitude of 70%). The THz field was then focused for detection a using 2"-diameter, 2"- focal distance off-axis mirror. To block the residual OPA beam after the generation crystal, we used 3 low pass filters (LPF) with cut off frequency of 18 THz with an out-of-band blocking is better than 0.1% each.

**Spectral features of the generated THz pulses.** The THz spectra produced by optical rectification in nonlinear crystals depend on the effective generation crystal length and the pump spectrum. For a transform-limited 65 fs pump pulse (FWHM)), the ideal THz spectrum should reflect an optically rectified spectrum centered around 5.7 THz with a FWHM of 9.5 THz [30]. However, DSTMS and OH1 (as most of the THz generating crystals) have strong phonon absorptions in this range that significantly modulate the spectrum. The amplitude spectra of THz pulses produced in DSTMS and OH1 are shown in Fig. 1b. The strong features in the spectra reflect the phonon resonance in this region. DSTMS has a strong absorption resonance at 1.024 THz. In the 1-4 THz range, there are no strong resonances [25]. We did not find data on resonances / linear characterization beyond 4 THz, but we observed a strong absorption around 4.9 THz, which likely comes from another resonance (Fig. 1). OH1 shows a higher spectral density at lower frequency going along with a wider spread of spectral density at least up to 12 THz. The main absorption resonances occur at 1.45 THz and 2.85 THz (consistent with previous reports [26]). Here again we could not find reports on higher frequencies resonances, but from our spectrum we anticipate resonances around {4.9, 6.3} THz. In the section "output pulse characteristics", we drew a comparison with ref [20]. Note that although the f-number of the detector focusing mirror in our measurement is different from ref [20], the numerical aperture (NA) of the main spectral contents was nearly the same as discussed in the article.

**Spot size imaging and calibration.** To image the THz spot size, we used a microbolometer camera from NEC. It is currently the most sensitive THz sensor on the market. However, the manufacturer states that the frequency range is 1-7 THz and is based on evaluation and calculated results with no guarantee due to the lack of a standard calibration method. We are unaware of any report or its use to measure a spot size < 270 µm nor of it being used at such high frequencies. In the estimation of the peak field from the measured energy, we considered the energy and the pulse duration of the main lobe (half cycle with peak field) of the pulse. Intensity calculations were performed following ref [31].

**Terahertz detection**. Throughout this work, we used two different detection s schemes: ABCD and EOS. A comparison between the measured pulses from the two techniques is shown in Fig. S1.

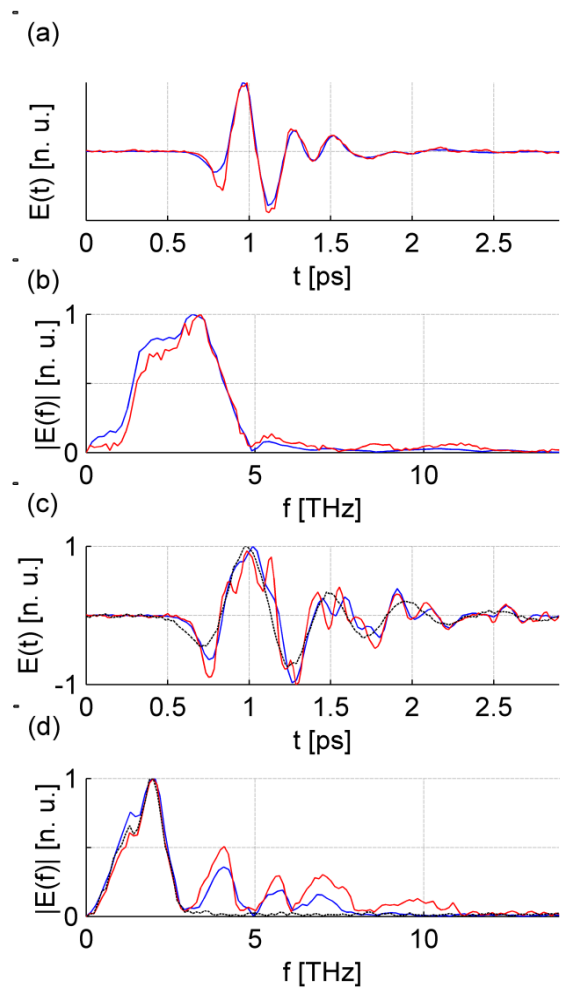

**Figure S1: Terahertz detection schemes.** A comparison between the measured THz from DSTMS ((a) & (b)) and OH1 ((c) & (d)). The blue curves show the measurement using EOS. ABCD measurements are shown in red. In the case of OH1, another measurement (black) was taken with ABCD after filtering out the higher frequency (> 3 THz) components.